\pdfoutput=1
\documentclass[acmsmall]{acmart}

\AtBeginDocument{%
  }

\setcopyright{cc}
\setcctype{by}
\acmDOI{10.1145/3715727}
\acmYear{2025}
\acmJournal{PACMSE}
\acmVolume{2}
\acmNumber{FSE}
\acmArticle{FSE012}
\acmMonth{7}
\received{2024-09-13}
\received[accepted]{2025-01-14}

\usepackage{framed}
\usepackage{listings}
\usepackage{xcolor}
\usepackage{color}
\usepackage{multirow}
\usepackage{algorithm}

\usepackage{amsmath}
\usepackage{amssymb}
\usepackage{threeparttable}
\usepackage{url}
\usepackage[most]{tcolorbox}
\usepackage{hyperref}
\usepackage[hyphenbreaks]{breakurl}
\usepackage{algpseudocode}
\usepackage{colortbl}
\usepackage{booktabs}
\usepackage{balance}
\hypersetup{
    colorlinks = true,
    linkcolor=blue,
    filecolor=blue,      
    urlcolor=blue,
    citecolor=cyan,
}
\usepackage{xspace}
\usepackage{subfigure}
\usepackage{graphicx}
\usepackage{fvextra}
\usepackage{pifont}

\usepackage[shortlabels]{enumitem}
\setlist[enumerate]{nosep}

\definecolor{codegreen}{rgb}{0,0.6,0}
\definecolor{codegray}{rgb}{0.5,0.5,0.5}
\definecolor{codepurple}{rgb}{0.58,0,0.82}
\definecolor{backcolour}{rgb}{0.95,0.95,0.92}
\definecolor{lightgreen}{rgb}{0,0.4,0}
\definecolor{lightred}{rgb}{0.4,0,0}

\definecolor{mygray}{gray}{.9}

\lstdefinestyle{mystyle}{
    commentstyle=\color{brown},
    keywordstyle=\color{magenta},
    numberstyle=\tiny\color{codegray},
    stringstyle=\color{codepurple},
    basicstyle=\ttfamily\footnotesize,
    breakatwhitespace=false,         
    breaklines=true,                 
    captionpos=b,                    
    keepspaces=true,                 
    numbers=none,                    
    numbersep=5pt,                  
    showspaces=false,                
    showstringspaces=false,
    showtabs=false,                  
    tabsize=2,
    escapeinside={<@}{@>}
}

\newcommand{\tool}{\textsc{STGen}\xspace}
\newcommand{\bench}{\textsc{COFFE}\xspace}

\newcommand\etal{{\it{et al.\ }}}

\newcommand{\eat}[1]{\if 0 #1 \fi}

\newcommand{\g}{\cellcolor{mygray}}

\newfont{\mycrnotice}{ptmr8t at 7pt}
\newfont{\myconfname}{ptmri8t at 7pt}

\usepackage[]{collab}
\collabAuthor{yc}{orange}{Yichen Li}

\newcommand{\answer}[2]{
\begin{tcolorbox}[breakable,width=\linewidth,boxrule=0pt,top=1pt, bottom=1pt, left=1pt,right=1pt, colback=gray!20,colframe=gray!20]
\textbf{Answer to RQ#1:} #2
\end{tcolorbox}
}

\newcommand{\finding}[2]{
\begin{tcolorbox}[breakable,width=\linewidth,boxrule=0pt,top=1pt, bottom=1pt, left=1pt,right=1pt, colback=gray!20,colframe=gray!20]
\textbf{Finding #1:} #2
\end{tcolorbox}
}

\begin{document}

\title{\bench: A \underline{Co}de E\underline{ff}ici\underline{e}ncy Benchmark for Code Generation}



\author{Yun Peng}
\orcid{0000-0003-1936-5598}
\affiliation{%
  \institution{The Chinese University of Hong Kong}
  \city{Hong Kong}
  \country{China}
}
\email{ypeng@cse.cuhk.edu.hk}

\author{Jun Wan}
\orcid{0009-0006-3294-688X}
\affiliation{%
  \institution{Zhejiang University}
  \city{Hangzhou}
  \country{China}
}
\email{22451014@zju.edu.cn}

\author{Yichen Li}
\orcid{0009-0009-8370-644X}
\affiliation{%
  \institution{The Chinese University of Hong Kong}
  \city{Hong Kong}
  \country{China}
}
\email{ycli21@cse.cuhk.edu.hk}

\author{Xiaoxue Ren}
\authornote{Corresponding author.}
\authornote{Also with Hangzhou High-Tech Zone (Binjiang) Institute of Blockchain and Data Security.}
\orcid{0000-0002-5526-1617}
\affiliation{%
  \institution{The State Key Laboratory of Blockchain and Data Security, Zhejiang University}
  \city{Hangzhou}
  \country{China}
}
\email{xxren@zju.edu.cn}




\begin{abstract}
Code generation has largely improved development efficiency in the era of large language models (LLMs). With the ability to follow instructions, current LLMs can be prompted to generate code solutions given detailed descriptions in natural language. Many research efforts are being devoted to improving the correctness of LLM-generated code, and many benchmarks are proposed to evaluate the correctness comprehensively. Despite the focus on correctness, the time efficiency of LLM-generated code solutions is under-explored. Current correctness benchmarks are not suitable for time efficiency evaluation since their test cases cannot well distinguish the time efficiency of different code solutions. Besides, the current execution time measurement is not stable and comprehensive, threatening the validity of the time efficiency evaluation.

To address the challenges in the time efficiency evaluation of code generation, we propose \bench, a code generation benchmark for evaluating the time efficiency of LLM-generated code solutions. \bench contains 398 and 358 problems for function-level and file-level code generation, respectively. To improve the distinguishability, we design a novel stressful test case generation approach with contracts and two new formats of test cases to improve the accuracy of generation. For the time evaluation metric, we propose efficienct@k based on CPU instruction count to ensure a stable and solid comparison between different solutions. We evaluate 14 popular LLMs on \bench and identify four findings. Based on the findings, we draw some implications for LLM researchers and software practitioners to facilitate future research and usage of LLMs in code generation.
\end{abstract}



\maketitle

\section{Introduction}\label{sec:intro}

Nowadays, large language models (LLMs) such as GPT-4~\cite{gpt4} and Llama3.1~\cite{llama31} have demonstrated great ability to solve different software engineering tasks. With the ability to follow instructions~\cite{chung22scaling,wei22emergent,ouyang22training,Muennighoff24octopack}, LLMs can act like human developers, promptly handle the instructions and generate completed code, reviews, or comments. Code generation, which is tasked with
converting natural language instructions into executable code, 
has the potential to significantly enhance the efficiency of software development.
It is thus a critical software engineering problem being studied by many researchers. Researchers have proposed different approaches to make use of LLMs on code generation via prompting engineering~\cite{shinn23reflexion,zhou23language,chen23teaching,ni23lever,ridnik24code}, multiple-agent cooperation~\cite{huang23agentcoder,hong24metagpt,wang24executable,zhang24autocoderover,yang24sweagent,holt23l2mac}, and retrieval augmentation~\cite{liu21retrieval,parvez21retrieval,zhou23docpromting,zhang23repocoder,su24arks}. 

To facilitate the evaluation of code generation, many benchmarks such as HumanEval~\cite{humaneval}, MBPP~\cite{mbpp}, CodeContests~\cite{codecontests}, and APPS~\cite{apps} are proposed to evaluate the correctness of generated code solutions, and we refer them as correctness benchmarks. These benchmarks include coding tasks drafted by experienced developers~\cite{humaneval,mbpp} or collected from coding competitions~\cite{codecontests,apps}, with several test cases for each problem to examine the correctness of LLM-generated code solutions. With the correctness benchmarks, researchers can thoroughly study and further improve the ability of LLMs to generate correct code. Built upon current advanced techniques, powerful LLMs such as GPT-4 have obtained remarkable performance with the Pass@1 of 86.6\% on the function-level code generation benchmark HumanEval~\cite{humaneval}, reported by the EvalPlus leaderboard~\cite{evalplusboard}.

However, correctness benchmarks alone are insufficient to comprehensively evaluate LLMs' ability of code generation, especially when these models are increasingly used to generate code solutions for software products~\cite{yu2024where}.  In real-world software development, both correctness and time efficiency are crucial for ensuring software quality. Correct but time-inefficient code can lead to a lot of CWE issues~\cite{cwe}. Recent work~\cite{shypula2024learning,peng2024perfcodegen,liu2024evaluating,liu2024learning} on LLM-based code generation steps further to generate correct and efficient code. They directly adopt existing correctness benchmarks and measure the execution time of LLM-generated code solutions to determine the time efficiency. We argue that current correctness benchmarks are not suitable for time efficiency evaluation for the following challenges:
\begin{itemize}[wide=0pt]
    \item \textbf{Challenge 1: Existing correctness test cases cannot well distinguish the time efficiency of different code solutions.} Test cases in correctness benchmarks usually have small inputs since they aim to cover most corner cases to detect potential logical errors in code solutions. However, such test cases can hardly distinguish the time efficiency of different code solutions since code with different time complexities may cost similar time under small inputs. Therefore, it is necessary to include test cases with larger inputs so that we can better distinguish code solutions with different time efficiency. We refer to such test cases as stressful test cases. Stressful test case generation is not straightforward and cannot be easily handled by current correctness test case generation methods. Stressful test cases usually consume much more execution time, so traditional execution-based test case generation methods with many iterations of complete executions are too time-consuming to be adopted. LLM-based test case generation methods without execution can generate stressful test cases quickly, but they are limited by context windows and can hardly maintain the long inputs in results, threatening the accuracy of stressful test case generation.
    \item \textbf{Challenge 2: Execution time metric is unstable and not comprehensive for time efficiency evaluation.}  Unlike correctness evaluation, which can be easily repeated on any computer machine, execution time measurements highly rely on the machine where the experiments are conducted. Shypula \etal~\cite{shypula2024learning} find that two single time measurements of the code solution on the same environment can differ as much as 1.91$\times$. Unstable execution time measurements threaten the validity of time efficiency evaluation. Besides, previous work~\cite{shypula2024learning,effibench} regards time efficiency evaluation as independent of correctness evaluation for code generation, but time efficiency evaluation is conducted upon correctly generated code solutions. Using separate metrics to evaluate the correctness and time efficiency makes it hard to distinguish the quality of code solutions with high correctness but low time efficiency and those with high time efficiency but low correctness. Currently, there is no single metric evaluating both the correctness and time efficiency of LLM-generated code solutions.

\end{itemize}

To address the two challenges above in evaluating the time efficiency of LLM-generated code solutions, we \textbf{1) propose a new time efficiency benchmark named \bench, along with a novel approach \tool to generate stressful test cases automatically}. 
Specifically, \bench is built upon existing correctness benchmarks HumanEval~\cite{humaneval}, MBPP~\cite{mbpp} for function-level code generation and CodeContests~\cite{codecontests}, and APPS~\cite{apps} for file-level code generation by adding stressful test cases generated by \tool. Hence, it contains two splits for function-level and file-level code generation. \tool implements three phases to improve the accuracy of stressful test case generation. In the first phase, \tool generates contracts that record the dependencies between inputs, and contracts are then used to guide the test case generation in the second phase. An LLM judge checks conflicts between generated contracts and test cases and rejects incorrect test cases in the third phase. By validating test cases on contracts, \tool can identify incorrect test cases early and provide feedback for LLMs to help fix them. \tool also uses expressions and generator functions to replace the raw inputs in the stressful test cases to avoid overlong test cases that hinder the generation of LLMs. 
Furthermore, we \textbf{2) propose a new metric named \textit{efficient@k} that considers both correctness and time efficiency based on \textit{CPU instruction count} measurements.}. Efficient@k follows the same logic as pass@k~\cite{humaneval}, and the difference is that it requires a code solution to be correct and faster than the best ground truth solution to contribute. When comparing code solutions and ground truth solutions, we replace execution time with a more stable measurement \textit{CPU instruction count} to conduct a solid comparison.

Experiments demonstrate that \tool is quite effective in stressful test case generation by correctly generating approximately 99\% of test cases with a 96\% line coverage. Furthermore, 
To evaluate the effectiveness of stressful test cases generated by \tool, the stressful test cases generated by \tool can much better distinguish the time efficiency of code solutions by achieving the relative standard deviation (RSD) of 27.26\% and 17.60\% over different function-level and file-level code solutions generated by Llama3.1~\cite{llama31}, largely improving the RSD of 19.05\% and 15.73\% on the original correctness test cases. This indicates the high quality of \bench. To verify the stability of CPU instruction count, we compare it with execution time and find that CPU instruction count has a RSD of 0.003\%$\sim$0.005\%, which is 1,000$\times$ smaller than that of execution time measurement (2.37\%$\sim$5.65\%). This provides a solid basis for the calculation of efficient@k.

Based on \bench, we evaluate the time efficiency of code solutions generated by ten open-source LLMs and four closed-source LLMs and identify the following important findings:
\begin{itemize}
    \item The performance of current LLMs drops significantly in efficient code generation, indicating that the code solutions generated by current LLMs are correct but not time-efficient.
    \item Compared with function-level code generation, code solutions generated by current LLMs are less efficient in file-level code generation.
    \item Larger LLMs generally perform better in correct code generation but do not significantly outperform smaller LLMs in efficient code generation, indicating larger parameter sizes of current LLMs do not contribute much to efficient code generation.
\end{itemize}

We summarize the contributions of this paper as follows:
\begin{itemize}
    \item We build \bench, a benchmark for evaluating the time efficiency of both function-level and file-level code solutions generated by LLMs.
    \item We propose \tool, the first LLM-based stressful test case generation approach that employs contract validation and test cases with expression and generator functions inputs to improve accuracy.
    \item We introduce a novel metric efficient@k, based on stable CPU instruction count measurement, to evaluate the correctness and time efficiency of the LLM-generated code solutions.
    \item We conduct extensive experiments to evaluate the quality of \bench, the effectiveness of \tool, and the ability of current LLMs to generate efficient code.
\end{itemize}

\section{Problem Definition}\label{sec:mot}

\begin{figure}[t]
    \centering
    \includegraphics[width = 1.0\textwidth]{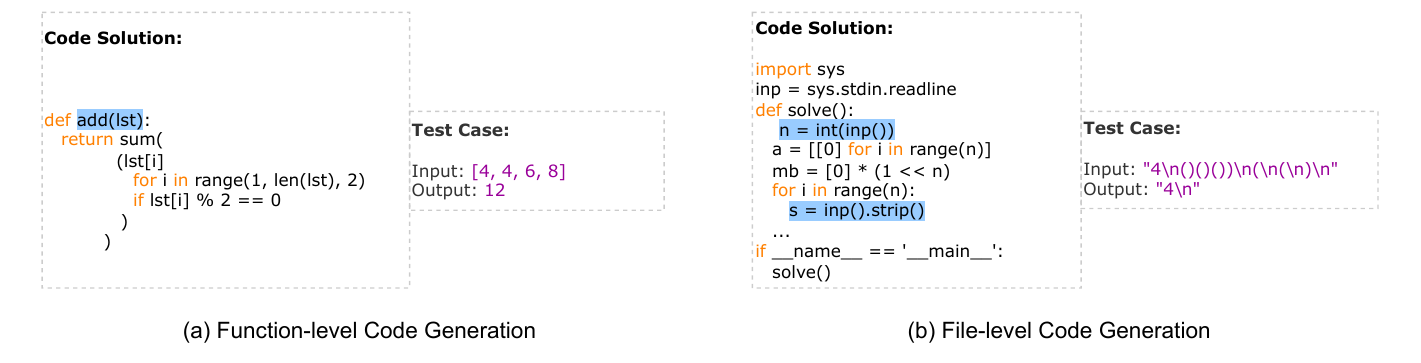}
    \caption{Examples for function-level and file-level code generation.}
    \label{fig:probdef}
\end{figure}

Currently, there are three types of code generation tasks: function-level, file-level, and repo-level code generation. We mainly focus on the first two types of code generation since repo-level code generation involves different modules in the repositories and third-party dependencies, making it hard to obtain solid time efficiency measurements. To better illustrate the differences between function-level and file-level code generation, we present two examples in Figure~\ref{fig:probdef}.

\textbf{Function-level Code Generation.} Function-level code generation takes natural language functionality descriptions as input and generates a single function that satisfies the requirements.
The generated function accepts inputs through function parameters.
The HumanEval~\cite{humaneval} and MBPP~\cite{mbpp} benchmarks are designed to benchmark function-level code generation.


Figure~\ref{fig:probdef}(a) shows an example function.
We observe that the function \textit{add()} only has a parameter named \texttt{lst}, and we only need to generate test inputs for this parameter to build a test case.
This shows that the number of parameters in functions is determined and functions accept inputs only once from parameters before the function execution.
Therefore, \textbf{to generate test cases for function-level code generation, we can generate test inputs for each parameter and combine them as a test case.}


\textbf{File-level Code Generation.} File-level code generation generates a complete program file instead of a single function to satisfy specified requirements. 
The inputs of the program file are managed by \textit{standard input (stdin)} related APIs, e.g., \textit{input()}.
File-level code generation tasks frequently appear in coding competitions, based on which researchers built Code Contests~\cite{codecontests} and APPS~\cite{apps} benchmarks.


Figure~\ref{fig:probdef}(b) shows an example program file.
We observe that this code solution accepts inputs in two locations (highlighted in blue). The input in the first location is used to control how many times the input in the second location will take.
This indicates that \textbf{the number of inputs for program files is not only determined by the code solution but also by the inputs.}
This poses great challenges in generating test cases for file-level code generation.


\section{Methodology}\label{sec:meth}

This section describes how we build the benchmark \bench, including selecting the coding problems, proposing \tool to generate stressful test cases for function-level and file-level code generation, and designing a novel time efficiency metric efficient@k.

\subsection{Data Preparation}\label{sec:data}

To construct \bench, we collect problems in the test splits of two existing function-level correctness benchmarks (i.e., HumanEval~\cite{humaneval} and MBPP~\cite{mbpp}), and two existing file-level correctness benchmarks (i.e., APPS~\cite{apps} and CodeContests~\cite{codecontests}). Each benchmark contains multiple coding problems and provides each problem with a description that explains the requirements in natural language, several ground truth solutions that address the problem, and several test cases that evaluate the correctness of generated code solutions. As there are multiple versions for MBPP, we choose the common subset of the sanitized version~\cite{mbppsantized} and the MBPP+ benchmark verified by EvalPlus~\cite{mbppplus} as our base benchmark to ensure the highest quality. 

With the selected benchmarks, we first validate the problems by checking the potential conflicts of provided test cases and ground truth solutions. Secondly, we select problems that most LLMs could correctly answer to reduce the difficulty of problems for the two file-level benchmarks since a problem is not useful in time efficiency evaluation if no LLM can answer it. We show the statistics of four benchmarks in Table~\ref{tab:datasets}.

\begin{table}[t]
    \centering
    \caption{The statistics of four sanitized benchmarks we selected to build \bench. ``Ori.'', ``Val.'' and ``Sel.'' indicate the original problems, validated problems, and finally selected problems in the benchmarks. The other columns in the table represent the data for the finally selected problems.}
    \scalebox{0.8}{
    \begin{tabular}{ccccccc}
    \toprule
        \multirow{2}*{\textbf{Benchmark}} & \multicolumn{3}{c}{\textbf{\#Problem}} & \multirow{2}*{\textbf{\#Solution/Problem}} & \multirow{2}*{\textbf{\#Test Case/Problem}} & \multirow{2}*{\textbf{Level}} \\
        \cmidrule{2-4}
        & Ori. & Val. & Sel. & & & \\
    \midrule
       HumanEval  & 164 & 164 & 164 & 1.00 & 9.57 & Function \\
       MBPP & 234 & 234 & 234 & 1.00 & 3.02 & Function \\
       Code Contests & 111 &  106 & 58 & 80.26 & 197.53 & File \\
       APPS & 5,000 & 3,106 & 300 & 64.36 & 13.94 & File \\
    \bottomrule
    \end{tabular}}
    \label{tab:datasets}
\end{table}

\subsubsection{Problem Validation}
To ensure the quality of test cases and ground truth solutions in the four benchmarks, we run the ground truth solutions in the provided test cases and remove 1) ground truth solutions that cannot pass the provided test cases to ensure consistency, 2) ground truth solutions with file operations to keep safety, and 3) problems without valid ground truth solutions and test cases. We show the number of validated problems in each benchmark in the third column of Table~\ref{tab:datasets}. All problems in HumanEval and MBPP can be successfully validated, so no problem is removed. For the Code Contests benchmark, we identify five problems with file operations, and we remove them to guarantee the safety of testing environments. For the APPS benchmark, we identify 1,894 problems whose ground truth solutions conflict with the provided test cases. The reason for such conflicts is that the APPS benchmark does not require the output of a code solution to exactly match the expected outputs in test cases to be correct, which differs from the other three benchmarks. We remove the 1,894 problems without exact matches in the APPS benchmark to maintain consistent evaluation standards.

\subsubsection{Problem Selection}
Current LLMs are quite effective in function-level code generation by achieving a pass@1 of more than 80\% in the HumanEval benchmark, as discussed in Sec.~\ref{sec:intro}. However, they perform much worse in file-level code generation since the most powerful LLM has a Pass@1 of 28.5\% on the Code Contests benchmark and a Pass@1 of less than 10\% on the APPS benchmark~\cite{codecontestleaderboard,appspsc}. This limits the usage of the full set of the Code Contests and APPS benchmarks because a problem that no LLM can correctly answer does not contribute to the time efficiency evaluation.
Therefore, for the validated problems in the two benchmarks, we sample one code solution with temperature 0 on 14 LLMs used in our experiments described in Table~\ref{tab:models} and remove 48 and 2,223 problems that code solutions from all LLMs failed in the Code Contests and APPS benchmark, respectively. To balance the number of problems in the function-level split and file-level split of \bench, we further select 300 problems in the APPS benchmark for which more LLMs can generate correct code solutions. We show the number of selected problems from the four benchmarks and associated statistics in the 4$\sim$6 columns of Table~\ref{tab:datasets}.

\subsection{Stressful Test Case Generation: \tool}
\begin{figure*}[tbp]
    \centering
    \includegraphics[width=1.0\textwidth]{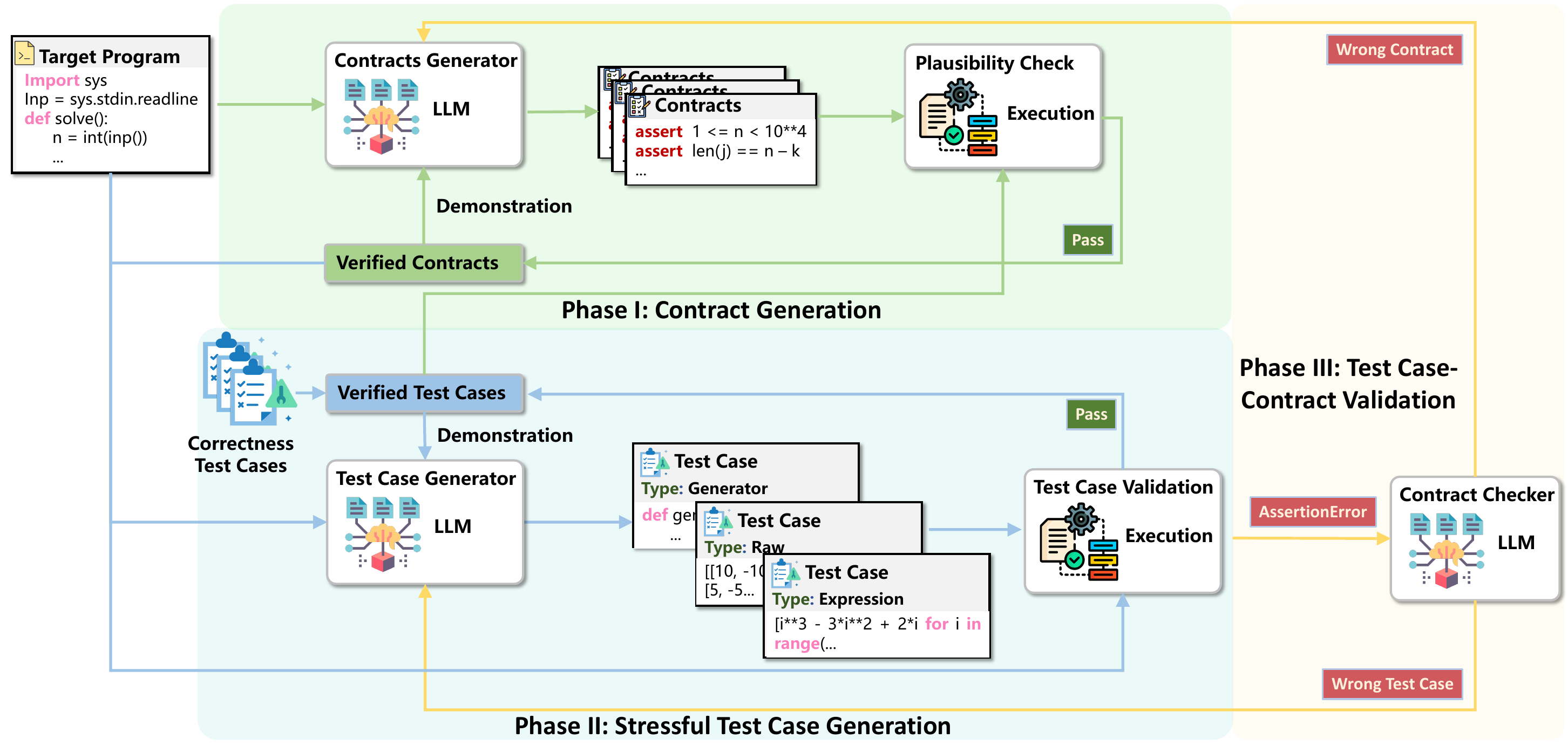}
    \vspace{-20pt}
    \caption{The overview workflow of \tool.} 
    \label{fig: overview}
    \vspace{-10pt}
\end{figure*}

With the selected problems, we propose a novel LLM-based approach \tool to generate stressful test cases automatically. In contrast to current LLM-based test case generation methods~\cite{liu2023is, li24large,ouédraogo24large,bhatia23unit,max24an,sami24a}, \tool aims to generate test cases to evaluate the time efficiency of code solutions under extreme conditions rigorously. This inherently requires constructing exceptionally long and intricate inputs that can hardly be handled by LLMs directly, leading to unsatisfactory accuracy, i.e., the proportion of correctly generated stressful test cases is low.

\subsubsection{Overview}

To improve the accuracy of stressful test case generation, \tool introduces contracts to guide the test case generation and validate the generated test cases. Contracts are collections of assertion statements that record the type, scale, and internal constraints between the inputs. Providing contracts in the test case generation process can help LLMs understand the dependencies between test inputs. Besides, \tool can easily identify incorrect test cases from the assertion errors contracts raise. To avoid overlong stressful test cases that hinder the performance of LLMs, we design two new formats of test cases by reformulating the test case generation task into a code generation task: \textit{expression test cases} and \textit{generator test cases}. Different from raw test cases that directly provide test inputs, expression and generator test cases contain code to generate test inputs, which greatly shortens the length of test cases.

We present the overview of \tool in Figure \ref{fig: overview}. \tool does not directly generate stressful test cases. Instead, it decomposes the task into three phases: 1) contract generation, 2) stressful test case generation, and 3) test case-contract pair check. In the first phase, \tool generates contracts by analyzing the target program, i.e., the ground truth solution for each problem in the benchmark. The generated contracts are then provided as demonstrations for stressful test case generation in the second phase, in which \tool generates expression and generator test cases instead of raw test cases. Since contracts are also generated and there is no guarantee of their correctness, \tool enters the third phase if the number of \textit{AssertionError} occurrences for a certain contract exceeds a threshold. In the third phase, \tool implements an LLM judge to determine the responsibility for conflicts between generated contracts and test cases. The contracts or test cases that are judged to be incorrect will be sent back for regeneration. This iterative process allows the generation of contracts and stressful test cases to mutually reinforce each other.

\subsubsection{Phase I: Contract Generation}\label{sec:contract}
In the first phase, \tool inserts assertion statements that check the preconditions of inputs as contracts into the target program, such as \textit{assert n > 0}. The contracts ensure that the inputs meet the required specifications in format (e.g., variable type), scale (e.g., input length, order of magnitude), and intrinsic constraints (e.g., right triangle side lengths).

The benefits of inserting contracts before stressful test case generation are twofold: 1) \textbf{Knowledge Enrichment.} Contracts explicitly indicate the functionality of the target program and the dependencies between inputs, which can help LLM better understand natural language descriptions provided in problems~\cite{liu2023is,endres2024can}; 2) \textbf{Early Validation.} Contracts can identify invalid inputs in test cases at the beginning of program execution and stop the execution-based test case validation process early, which largely improves the efficiency of the test case generation process.

In contract generation, \tool generates one assertion statement in an iteration and combines all assertion statements into a contract. When generating assertion statements, \tool prompts LLMs to consider the type, scale, and intrinsic constraints between inputs given the target program, existing correctness test cases, and previously generated assertion statements as demonstrations. \tool implements the same methodology to generate assertion statements for function-level and file-level target programs. However, \tool employs different strategies to insert assertion statements into target programs, given the differences between the code solutions in function-level and file-level code generation illustrated in Sec.~\ref{sec:mot}.

\textbf{Function-level Contract Insertion.} For function-level target programs with a determined number of inputs, \tool generates and inserts assertion statements for function parameters at the beginning of the function body. For example, \tool inserts assertion statements right before the return statements in the function \textit{add()} in Figure ~\ref{fig:probdef}(a).

\textbf{File-level Contract Insertion.} For file-level target programs with an unknown number of inputs and multiple input locations, \tool reformulates the contract generation problem into a code editing problem. It first identifies all input locations by checking the related system APIs such as \textit{input()} and then inserts assertion statements for each identified input location sequentially. \tool inserts assertion statements right after the input locations in most cases. However, as input locations in loops generally assign values for generic types such as \textit{list} and \textit{dict}, \tool inserts assertion statements after the entire loop where the assignments are complete to check the fully assigned types. For example, \tool identifies two input locations highlighted in blue in Figure~\ref{fig:probdef}(b). \tool first generates assertion statements for the input that assigns values to variable $n$ and inserts them right after the assignment. \tool then generates assertion statements for the second input in the loop, and this time, it inserts assertion statements after the entire \textit{for} loop.

To improve the correctness of generated assertion statements, \tool tests all generated assertion statements against the correctness test cases each time it inserts a new assertion statement. If the current assertion statement fails on the test cases, \tool only removes the current assertion statement and regenerates a new one while maintaining the assertion statements correctly generated in previous iterations. The iteration ends until no new assertion statements are generated or a maximum iteration number is reached.

\subsubsection{Phase II: Stressful Test Case Generation}
With the generated contracts as demonstrations, in the second phase, \tool generates stressful test cases. Unlike correctness test case generation, it is quite challenging to generate stressful test cases because LLMs must generate test cases of maximal length and complexity within the constraints of its finite context window while simultaneously ensuring adherence to intrinsic input constraints specified by contracts. Correctness test cases in current benchmarks~\cite{humaneval,mbpp,codecontests,apps} are raw test cases that directly provide the values for test inputs. However, due to the limited context window size, it is infeasible to directly generate overlong raw test cases for time efficiency evaluation. For example, it is hard for LLMs to generate a list with more than a million numbers for stressful tests. To address this challenge, we introduce two new formats of stressful test cases:

\textbf{Expression Test Cases.} Expression test cases utilize Python expressions to generate test cases, allowing for more complex input generation while maintaining a compact representation within the LLM's context window. For instance, a list with a million numbers could be easily generated by an expression ``\textit{[random.randint(1, 100000) for \_ in range(1000000)]}'', which is much shorter than listing a million numbers. Expression test cases offer a balance between complexity and conciseness, enabling the creation of structured inputs. They are suitable for function-level test case generation with a determined number of test inputs. To evaluate code solutions on expression test cases, we just need to execute the expressions to get the real test inputs before the code execution.

\textbf{Generator Test Cases.} Generator test cases are Python functions that output the test inputs. It is quite useful for creating stressful test cases that require intricate logical relationships or patterns that are difficult to express in single expressions. For example, it is suitable for file-level code generation where the number of inputs is undetermined. Expression test cases cannot handle this since we do not know how many expressions should be generated.

To generate expression and generator test cases, \tool prompts the LLMs with contract, verified generated test cases as demonstrations, so LLMs can learn the dependencies between inputs as well as the specific formats of the expected test cases. The generated test cases are then verified against the previously generated contracts and the target program. Test cases that pass the validation of contracts and the execution of the target program are collected to build \bench. Verified stressful test cases are also used as demonstrations to help generate the following stressful test cases.

\subsubsection{Phase III: Test Case-Contract Pair Check}

Although the generated contracts are verified against the existing correctness test cases, correctness test cases do not cover all possible cases and dependencies among inputs, especially in stressful scenarios. Contracts can still make mistakes and induce false positives. During the test case validation, if a generated test case violates the inserted contract, it triggers an \textit{AssertionError}. If the \textit{AssertionError} consistently occurs for multiple test cases, the contract may be incorrect and thereby hinder the entire stressful test case generation procedure. To mitigate this, when the number of conflicts between contracts and test cases (i.e., \textit{AssertionError} occurrences during execution) exceeds a predefined threshold (5 in this paper), the generated test case and the violated contract are paired for further check by an LLM judge checker in the third phase.

The LLM judge takes all accumulated contract-related execution failure pairs as inputs, along with the target program, to analyze and determine the validity of the contracts and the test cases. The judge reviews the violated contract with exact stressful inputs, rethinks the correctness of the generated contract, and determines the root cause of conflicts. Once the root cause is identified, the relevant judgment results and corresponding failure pairs are sent back to the previous phases for regeneration. By providing feedback for incorrect contracts or test cases, \tool enhances the robustness of the test case validation and enables the improvements between contract generation and test case generation. To prevent duplicate judgments, once the LLM judge determines that a contract is valid in the third phase, it will not be checked again, and test cases that fail the validation of this contract will be directly rejected in the future.

\subsection{Time Efficiency Metric: Efficient@k}\label{sec:metric}

Previous work~\cite{shypula2024learning,effibench} intuitively adopts execution time as the performance measurement to evaluate the time efficiency of LLM-generated code. However, execution time measurements could be affected by many factors, such as process scheduling and disk I/O, so it is not stable enough to make a solid comparison between the time efficiency of different code solutions. In this section, we propose to use \textit{CPU instruction count} to replace execution time to measure the time efficiency of code solutions stably. Based on CPU instruction count measurements, we propose a new metric \textit{efficient@k} to evaluate both the correctness and time efficiency of code solutions.

\subsubsection{CPU Instruction Count} To find a more stable measurement to replace execution time, we first look into the factors contributing to the execution time. Patterson and Hennessy~\cite{arch} define the CPU time cost by a program through the following equation.
\begin{equation}\label{eq:cic}
    \text{CPU Time} = [\textbf{\text{Instruction Count}}] \times [\text{Clock per Instruction}] \times [\text{Clock Cycle Time}]
\end{equation}

From the equation, the CPU time of a program is determined by three factors. While \textit{Clock per Instruction} and \textit{Clock Cycle Time} depend on the physical machine where the program runs, the only factor related to the program is \textit{Instruction Count}. Therefore, if a program has a higher CPU instruction count on the same machine, it is less efficient, and vice versa. Unlike the execution time measurements that could be affected by many factors, CPU instruction count measurements are more stable as CPU instruction count for a program does not increase even if the program execution is slowed or stalled by external factors. It is also straightforward to measure CPU instruction count using the system APIs. For example, Linux provides a command tool named \textit{perf}~\cite{perf} to support CPU instruction count measurements.

\subsubsection{Efficient@k} CPU instruction count is a stable measurement for the time efficiency evaluation of different code solutions. However, its absolute value is not meaningful as the same code solution has different CPU instruction counts in different machines. Besides, it is not comprehensive as it does not measure the correctness of generated code solutions. To address these problems, we propose a new metric named \textit{Efficient@k}, inspired by the design of \textit{pass@k}~\cite{humaneval}. We show the original definition of pass@k in Equation~\ref{eq:pass} and the definition of proposed efficient@k in Equation~\ref{eq:eff}.

\begin{equation}\label{eq:pass}
    \text{pass@k}:=\underset{\text { Problems }}{\mathbb{E}}\left[1-\frac{\binom{n-c}{k}}{\binom{n}{k}}\right]
\end{equation}

\begin{equation}\label{eq:eff}
    \text{efficient@k}:=\underset{\text { Problems }}{\mathbb{E}}\left[1-\frac{\binom{n-c_f}{k}}{\binom{n}{k}}\right]
\end{equation}

Pass@k is an expectation over all problems in the benchmark for the probability that at least one solution in $k$ samples can pass all test cases. In equation~\ref{eq:pass}, total $n$ solutions are sampled from LLMs instead of only $k$ samples to reduce the variance. By running the sampled code solutions on correctness test cases, we can get the solutions $c$ that can pass all the test cases to estimate the probability of correctness. Pass@k is a solid metric with low variance and can be easily reproduced under different platforms. 

We follow the idea of pass@k when designing efficient@k. Pass@k requires the correct code solutions $c$ to contribute, while in efficient@k, we collect the number $c_f$ of the correct solutions faster than the best ground truth solution to replace $c$ in pass@k. Therefore, efficient@k evaluates the probability of LLMs to generate correct and fast enough code solutions. Efficient@k compares the CPU instruction count of code solutions and ground truth solutions to determine which runs faster. By doing so, efficient@k does not consider the absolute values of CPU instruction counts to avoid the impacts of specific systems or machines. With a value range from 0 to pass@k, efficient@k combines correctness and time efficiency evaluation to comprehensively evaluate the quality of code solutions.

\subsection{Code Efficiency Benchmark: \bench}

\begin{table}[t]
    \centering
    \caption{The statistics of \bench.}
    \scalebox{0.8}{
    \begin{tabular}{ccccc}
    \toprule
      \multirow{2}*{\textbf{Category}} & \multirow{2}*{\textbf{\#Problem}}  & \multirow{2}*{\textbf{\#Solution/Problem}}  & \multicolumn{2}{c}{\textbf{\#Test Case/Problem}} \\
      \cmidrule{4-5}
      & & & \textbf{Correctness} & \textbf{Stressful} \\
    \midrule
       Function-level  & 398 & 1.00 & 5.72 & 4.99 \\
       File-level & 358 &  66.93 & 43.68 & 4.95 \\
    \bottomrule
    \end{tabular}}
    \label{tab:benchmark}
\end{table}

With the stressful test case generation approach \tool, we add stressful test cases for each problem selected in Sec.~\ref{sec:data}. Specifically, we generate 20 stressful test cases for each problem and measure the CPU instruction count each test case costs. We conduct the measurements 12 times and remove the highest and lowest measurements before calculating the average to ensure the most stable results. In the CPU instruction measurements, we limit the execution time of one single measurement to five seconds so that the measurements for one test case will not exceed one minute. We then rank the average CPU instruction count of each test case and include the five test cases with the highest CPU instruction counts in \bench. We do not include all generated stressful test cases in \bench to avoid large time costs in time efficiency evaluation, since stressful test cases generally take much longer time than correctness test cases to execute. We reserve all existing correctness in \bench to validate the correctness of generated code solutions. We show the statistics of \bench in Table~\ref{tab:benchmark}.
\section{Experiment Setup}\label{sec:setup}

\subsection{Research Questions}
We focus on the following research questions:
\begin{itemize}
    \item \textbf{RQ1:} How well does CPU instruction count measure time efficiency compared with execution time?
    \item \textbf{RQ2:} How effective is \tool on stressful test case generation and how well are the generated stressful test cases?
    \item \textbf{RQ3:} How efficient is the code generated by current LLMs?
\end{itemize}

\subsection{Metrics}
To evaluate the stability of CPU instruction count (RQ1), we introduce the following metrics:
\begin{itemize}
    \item \textbf{Relative Standard Deviation (RSD):} The ratio of the standard deviation to the mean. We use it to measure how stable a performance metric is on the same code solution (the lower, the better) and how well a test case can distinguish different code solutions (the higher, the better). We use ``RSD (-)'' when it is used to evaluate stability and ``RSD (+)'' when it is used to evaluate distinguishability.
    \item \textbf{Pearson Correlation Coefficient:} The ratio between the covariance of two variables and the product of their standard deviations. We use it to measure the linear correlation between two metrics.
\end{itemize}

To evaluate the quality of stressful test cases and the effectiveness of \tool (RQ2), we introduce the following metrics:
\begin{itemize}
    \item \textbf{Accuracy:} The proportion of test cases generated by a certain method where the target program does not fail.
    \item \textbf{Line Coverage:} The percentage of executed lines in solutions when executing the test cases.
\end{itemize}

To evaluate the efficiency of code solutions generated by LLMs (RQ3), we use the following metrics:
\begin{itemize}
    \item \textbf{Pass@k:} The probability that at least one of the top k-generated code samples for a problem passes the unit tests, as illustrated in Sec.~\ref{sec:metric}.
    \item \textbf{Speedup:} The ratio $\frac{gt}{o}$ of CPU instruction count of best ground truth solution $gt$ to the CPU instruction count of a code solution $o$.
    \item \textbf{Efficient@k:} The probability that at least one of the top k-generated code samples for a problem is correct and more efficient than the best ground truth solution, as introduced in Sec.~\ref{sec:metric}.
\end{itemize}

\subsection{Baselines}
Since there is no previous work on LLM-based stressful test case generation, we select three widely used LLM-based correctness test case generation methods and adapt them into stressful test case generation:
\begin{itemize}
    \item \textbf{Instruction Prompting~\cite{wang24testeval}.} Wang \etal design several instruction prompt templates to ask LLMs to cover certain lines, branches, or paths of the code in test case generation. We modified their instruction prompt and let LLMs focus on stressful test case generation. This method generates raw test cases.
    \item \textbf{Few-shot Prompting~\cite{ouédraogo24large}.} Few-shot prompting adds several demonstrations to guide LLMs to generate similar test cases. This method generates raw test cases.
    \item \textbf{Generator-based Prompting~\cite{liu24llm}.} Instead of directly generating test cases, this method prompts LLMs to generate a function that derives the test cases. We adapt this method to our stressful test case generation and let LLMs generate functions that produce stressful test cases. This method generates generator test cases.
\end{itemize}

\subsection{Models}

To investigate the efficiency of code generated by current LLMs, we select 14 popular models for evaluation. We show the model names, sizes, and context lengths in Table~\ref{tab:models}. For GPT-3.5~\cite{chatgpt} and GPT-4o~\cite{gpt4o}, we use the APIs provided by OpenAI~\cite{openai} under engines ``\textit{gpt-3.5-turbo}'' and ``\textit{gpt-4o}'', respectively. For DeepSeek V2~\cite{deepseekv2} and DeepSeek V2 Coder~\cite{deepseekcoderv2}, we use the APIs provided by DeepSeek~\cite{deepseek} under engine ``\textit{DeepSeek-V2-0628}'' and ``\textit{DeepSeek-V2-0724}'', respectively. For Claude 3.5 Sonnect~\cite{claude}, we use the APIs provided by Anthropic~\cite{anthropic} under the engine ``\textit{claude-3-5-sonnet-20240620}''. For Gemini 1.5 Pro, we use the APIs provided by Google~\cite{googleapi} under the engine ``\textit{gemini-1.5-pro}'' to generate code solutions.   Due to limited computing resources, for open-source models larger than 13B, we use the API provided by Deep Infra~\cite{deepinfra} to generate code solutions. 

\begin{table}[t]
    \centering
    \caption{The LLMs we evaluate in this paper. Models highlighted in \colorbox{mygray}{gray} are closed-source models.}
    \scalebox{0.8}{
    \begin{tabular}{ccc|ccc}
    \toprule
        \textbf{Model} & \textbf{Size} & \textbf{Context Size} & \textbf{Model} & \textbf{Size} & \textbf{Context Size}\\
    \midrule
        Phi3~\cite{phi3} & 3.8B & 128k & MagicCoder~\cite{magiccoder} & DS-6.7B/CL-7B & 16,384\\
        CodeLlama~\cite{codellama} & 7B/13B/34B & 16,384 &
        Llama3~\cite{llama3} & 8B/70B & 4,096 \\
        StarCoder~\cite{starcoder} & 15B & 16,384 & WizardCoder~\cite{wizardcoder} & 15B & 2,048 \\
        Mixtral~\cite{mixtral} & 8$\times$7B & 32,768 &
        DeepSeek V2~\cite{deepseekv2} & 236B & 128k \\
        DeepSeek Coder V2~\cite{deepseekcoderv2}& 236B & 128k &
        Llama3.1~\cite{llama31} & 405B & 4,096 \\
        \g Claude 3.5 Sonnet~\cite{claude} & \g - & \g 200k & \g Gemini 1.5 Pro~\cite{gemini} & \g - & \g 200k \\
        \g GPT-3.5~\cite{mixtral} & \g - & \g 16,385 & \g GPT-4o~\cite{gpt4o} & \g - & \g 128k \\
    \bottomrule
    \end{tabular}}
    \label{tab:models}
\end{table}

\subsection{Implementation Details}

We conduct all experiments on a Linux machine with Ubuntu 20.04.4 LTS. It has an Intel(R) Xeon(R) Platinum 8358P CPU of 2.60G HZ with 128 cores and 2 TB memory. We use the Coverage.py~\cite{coverage} library to measure the line coverage of test cases, and the Cirron~\cite{cirron} library to measure the CPU instruction count a program consumes. 
\section{Experiment Results}\label{sec:eval}

\subsection{RQ1: CPU Instruction Count vs. Execution Time}

To demonstrate that CPU instruction count is more suitable for time efficiency evaluation than execution time, we focus on two aspects: stability, which evaluates how solid the measurement is, and correlation, which evaluates how close two measurements are.

\textbf{Stability.} To compare the stability of CPU instruction count and execution time measurements, we run the ground truth solutions of the validated problems on the correctness test cases from the four correctness benchmarks. Note that we do not run them on our stressful test cases to ensure a fair comparison since CPU instruction count is involved in building \bench. We run each solution 12 times and remove the largest and smallest measurements. We then calculate the RSD of the remaining 10 measurements and show the results in the second and third columns of Table~\ref{tab:metric}.

As the experiments are repeated on the same ground truth solution and same test cases, a lower relative standard deviation indicates a more stable measurement. From Table~\ref{tab:metric}, we can observe that execution time has an RSD of about 5\% on function-level benchmarks HumanEval and MBPP and an RSD of about 2\% on file-level benchmarks Code Contests and APPS. On the contrary, CPU instruction count has a more than 1000$\times$ smaller RSD (0.003\%$\sim$0.005\%) than execution time on four benchmarks. This indicates that the ten measurements of CPU instruction count almost remain the same on the same program, and CPU instruction count is quite stable in measuring time efficiency.

\textbf{Correlation.} To validate the linear correlation between CPU instruction count and execution time, as described in Equation~\ref{eq:cic}, we calculate the Pearson correlation coefficient between CPU instruction count and execution time, as shown in the last column of Table~\ref{tab:metric}. We find that the correlations of the two measurements on all benchmarks are very close to 1.0. This indicates that the two measurements are linearly correlated and verifies the correctness of Equation~\ref{eq:cic} as the other two factors \textit{Clock per Instruction} and \textit{Clock Cycle Time} do not change in the same testing environment. Therefore, we can replace execution time with CPU instruction count to measure time efficiency. 

\begin{table}[t]
    \centering
    \caption{Comparison between CPU Instruction Count and Execution time on different benchmarks. ``RSD (-)'' indicates the average relative standard deviation of a certain metric when running multiple times on the same ground truth solution. It evaluates the stability of different measurements. ``Correlation'' indicates the Pearson correlation coefficient between CPU instruction count and Execution time.}
    \scalebox{0.9}{
    \begin{tabular}{cccc}
    \toprule
        \multirow{2}*{\textbf{Benchmark}} & \multicolumn{2}{c}{\textbf{RSD (-)}} & \multirow{2}*{\textbf{Correlation}} \\
    \cmidrule{2-3}
        & \textbf{CPU Instruction Count} & \textbf{Execution Time} & \\
    \midrule
        HumanEval & 0.005\%  & 5.65\% & 1.00 \\
        MBPP & 0.004\% & 5.31\% & 1.00 \\
        Code Contests & 0.003\% & 2.37\% & 0.99 \\
        APPS & 0.003\% & 2.47\% & 0.96 \\
    \bottomrule
    \end{tabular}}
    \label{tab:metric}
\end{table}

\answer{1}{CPU instruction count is more suitable to evaluate time efficiency since it is much more stable than execution time by achieving a 1000$\times$ smaller RSD of 0.003\%$\sim$0.005\%, and it is linearly correlated with execution time with Pearson correlation coefficient of 0.96$\sim$1.0.}

\subsection{RQ2: Effectiveness of \tool and Distinguishability of Stressful Test Cases}
To answer RQ2, we study the effectiveness of \tool on stressful test case generation compared with three widely used LLM-based test case generation baselines. For the generated stressful test cases, we evaluate whether they can better distinguish different code solutions generated by LLMs. We show the main results of the comparison between \tool and baselines in Table~\ref{tab:testcase}.

\begin{table}[t]
    \centering
    \caption{Comparison between different test cases. ``Correctness'' indicates the original correctness test cases. ``Instruction'', ``Few-shot'' and ``Generator'' indicate the stressful test cases generated by three baselines, respectively.  ``RSD (+)'' indicates the relative standard deviation achieved by test cases on different code solutions generated by two powerful LLMs GPT-4o and Llama3.1. It evaluates the distinguishability of different test cases in terms of time efficiency. }
    \scalebox{0.85}{
    \begin{tabular}{cc|cc|cc}
    \toprule
     \multirow{2}*{\textbf{Level}}  & \multirow{2}*{\textbf{Method}} & \multirow{2}*{\textbf{Accuracy}} & \multirow{2}*{\textbf{Line Cov.}} & \multicolumn{2}{c}{\textbf{RSD (+)}}  \\
     \cmidrule{5-6}
      &  & &  &  \textbf{Llama3.1} & \textbf{GPT-4o} \\
    \midrule
      \multirow{5}*{Function} & Correctness &  - & 98.46 & 19.05\% & 21.35\% \\
      \cmidrule{2-6}
       & Instruction & 83.67  & 83.87 & 20.69\% & 22.21\% \\
       & Few-shot & 87.07 & 81.46 & 22.32\% & 20.61\% \\
       & Generator & 86.91 & 81.84 & 22.86\% & 21.32\% \\
       & \tool & \textbf{98.64} & \textbf{96.01} & \textbf{27.26\%} & \textbf{28.20\%}  \\
    \midrule
    \multirow{5}*{File} & Correctness & - & 95.68 &  15.73\% & 12.99\% \\
    \cmidrule{2-6}
       & Instruction & 84.52  & 85.29 &14.21\%  & 11.04\%  \\
       & Few-shot & 65.17 & 66.53 & 12.02\%   &  8.95\%  \\
       & Generator & 94.86 & 94.79 & 15.54\%  & 13.17\% \\
       & \tool & \textbf{98.91} & \textbf{95.17} & \textbf{17.60\%} & \textbf{14.79\%}  \\
    \bottomrule
    \end{tabular}}
    
    \label{tab:testcase}
\end{table}

\textbf{Effectiveness of \tool.} To study how contracts can improve the accuracy of test cases, we compare \tool with three baselines without contracts and report the accuracy at the third column of Table~\ref{tab:testcase} for function-level and file-level splits of \bench. We do not report the accuracy of the original test cases because they are manually drafted. From the table, we observe that \tool achieves an accuracy of 98.64\% and 98.91\%, outperforming the baselines by up to 17.89\% and 51.77\% in function-level and file-level splits, respectively. This suggests that almost all stressful test cases generated by \tool are correct. Without knowledge enrichment and early validation by contracts, on the contrary, baselines fail to generate about 5\%$\sim$35\% of stressful test cases. 

Apart from the accuracy, a correct test case is representative if it can cover most lines of the target program. To ensure the quality of generated stressful test cases, we evaluate the line coverage and report the results in the fourth column of Table~\ref{tab:testcase}. We find that \tool consistently achieves the highest line coverage of 96.01\% and 95.17\% for function-level and file-level stressful test cases, respectively. This demonstrates that the stressful test cases generated by \tool can thoroughly evaluate the time efficiency of the major code logic in target programs. We also note that the line coverage achieved by \tool is slightly lower than that achieved by the original correctness test cases. This is reasonable because the stressful test cases are much fewer than the correctness test cases in \bench, as can be seen in Table~\ref{tab:benchmark}.

\textbf{Distinguishability of Stressful Test Cases.} To evaluate how well the stressful test cases generated by \tool can distinguish the time efficiency of different code solutions, we sample 20 code solutions from two powerful LLMs, Llama3.1 and GPT-4o, for each problem in \bench. We then run the sampled solutions on different test cases and collect the CPU instruction count usage. We calculate the RSD on the CPU instruction counts of the sampled 20 code solutions, and a higher RSD indicates better distinguishability. We report the RSD on the code solutions of two models at the fifth and sixth columns of Table~\ref{tab:testcase}.  

Firstly, we observe that stressful test cases generated by \tool improve the RSD of original correctness test cases by 43.10\% and 32.08\% on Llama3.1 and GPT-4o, respectively, at the function level, and the improvements are 11.89\% and 13.86\% on Llama3.1 and GPT-4o, respectively, at the file level. \tool also outperforms all three baselines in terms of RSD on both Llama3.1 and GPT-4o. This demonstrates that stressful test cases generated by \tool can better distinguish different code solutions than original correctness test cases and stressful test cases generated by baselines. Secondly, we find that the generator-based prompting method achieves higher RSD than other baselines. This verifies the effectiveness of generator test cases compared with raw test cases in time efficiency evaluation. However, the generator-based prompting method cannot well handle multiple parameters in function-level programs by achieving an accuracy of only 86.91\%. \tool mitigates this problem by generating expression test cases that follow the formats of raw test cases but introduce small expressions to represent each input. As a result, the expression test cases generated by \tool for function-level code solutions outperform the generator-based prompting method by 19.25\% and 32.27\% in terms of RSD on Llama3.1 and GPT-4o, respectively.

\answer{2}{With knowledge enrichment and early validation by contracts, \tool is quite effective in generating correct stressful test cases with an accuracy of about 99\% and line coverage of about 96\%. The expression and generator test cases generated by \tool can better distinguish different code solutions' time efficiency with an RSD of up to 28.20\% on GPT-4o.}

\subsection{RQ3: Time Efficiency of Code Generated by LLMs}
Based on \bench, we evaluate the time efficiency of code generated by different LLMs. We select ten popular open-source LLMs and four popular closed-source LLMs, as shown in Table~\ref{tab:models}. We show the Pass@1, efficient@1, and speedup of all LLMs on \bench in Table~\ref{tab:modelres}.

\begin{table}[t]
    \centering
    \caption{The correctness and time efficiency of code solutions generated by LLMs in Table~\ref{tab:models} on \bench. Efficienct@1 and pass@1 are calculated upon all instances in \bench, and speedup is calculated on correct solutions generated by models. Models highlighted in \colorbox{mygray}{gray} are closed-source models. ``$\Delta$'' indicates the difference of efficient@1 and pass@1 in percentage (100\% - efficient@1 / pass@1).}
    \scalebox{0.8}{
    \begin{tabular}{cc|ccc|ccc}
    \toprule
        \multirow{2}*{\textbf{Model}}  & \multirow{2}*{\textbf{Size}} & \multicolumn{3}{c|}{\textbf{Function-level}} & \multicolumn{3}{c}{\textbf{File-level}} \\
        \cmidrule{3-5}
        \cmidrule{6-8}
        & & \textbf{Efficient@1 ($\Delta$)} & \textbf{Speedup} & \textbf{Pass@1} & \textbf{Efficient@1 ($\Delta$)} & \textbf{Speedup} & \textbf{Pass@1} \\
    \midrule
        Phi3 & 3.8B & 26.65 (39\%) & 2.59 & 43.47 & 7.36 (67\%) & 0.08 & 22.63 \\
    \midrule
        \multirow{2}*{MagicCoder} & DS-6.7B & 21.90 (32\%) & 3.04 & 32.41  & 12.02 (48\%) & 0.10 & 22.91 \\
        & CL-7B & 29.82 (36\%) & 3.41 & 46.48 & 5.04 (68\%) & 0.14 & 15.92 \\
    \midrule
        \multirow{3}*{CodeLlama} & 7B & 26.65 (31\%) & 2.49  & 38.69 & 4.26 (51\%) & 0.95 & 8.66 \\
        & 13B & 25.60 (39\%) & 1.03 &41.71  & 1.16 (48\%) & 1.02 & 2.23 \\
        & 34B & 40.37 (38\%) & 3.51 & 64.74 & 22.87 (57\%) & 0.09 & 53.63 \\
    \midrule
        \multirow{2}*{Llama3} & 8B & 27.70 (35\%)  & 3.91 & 42.46 & 0.00 (100\%) & 0.21 & 0.84\\
        & 70B & 40.90 (39\%) & 3.30 & 67.59 & 38.76 (44\%) & 0.14 & 68.99 \\
    \midrule
        StarCoder & 15B & 38.52 (37\%)  & 3.52 & 61.31 & 21.71 (58\%) & 0.10 & 51.11 \\
    \midrule
        WizardCoder & 15B & 28.76 (41\%) & 1.95 & 48.49 & 10.08 (51\%) & 0.07 & 20.67 \\
    \midrule
        Mixtral & 8$\times$7B & 25.59 (43\%) & 5.14 & 44.72 & 8.53 (63\%) & \textbf{1.43} & 22.91\\
    \midrule
        DeepSeek V2 & 236B & 46.70 (40\%) & 2.79 & 78.39 & 41.09 (54\%) & 0.18 & 89.94 \\
    \midrule
        DeepSeek V2 Coder & 236B &  \textbf{46.97} (41\%) & 2.53 & \textbf{79.90}  & 42.25 (46\%) & 0.44 & 78.77 \\
    \midrule
        Llama3.1 & 405B & 39.58 (41\%) & 3.21 & 67.34 & \textbf{46.51} (48\%) & 0.90 & 89.11 \\
    \midrule
        \g Claude 3.5 Sonnet & \g - & \g 43.54 (44\%) & \g 4.90 & \g 77.64 & \g 39.15 (55\%) & \g 0.23 & \g 86.59  \\
    \midrule
        \g Gemini 1.5 Pro & \g - & \g 45.12 (40\%) & \g 1.76 & \g 75.38 & \g 42.64 (43\%) & \g 0.16 & \g 75.44 \\
    \midrule
        \g ChatGPT & \g - & \g 37.73 (45\%) & \g 2.46 & \g 68.19 & \g 39.15 (48\%) & \g 0.12 & \g 75.98\\
    \midrule
        \g GPT-4o & \g - & \g 44.59 (43\%) & \g \textbf{8.28} & \g 77.64 & \g 43.02 (53\%)  & \g 1.11 & \g \textbf{90.78} \\
    \bottomrule
    \end{tabular}}
    
    \label{tab:modelres}
\end{table}

\textbf{Overall Time Efficiency.} 
To evaluate the time efficiency of code generated by different models, we study the efficient@1 and speedup. We use efficient@1 to evaluate the probability of an LLM to generate a correct code solution faster than the best ground truth solution and speedup to evaluate how fast the correctly generated code solutions are compared with the best ground truth solutions. From Table~\ref{tab:modelres}, we identify that DeepSeek V2 Coder obtains the highest efficient@1 of 46.97\% at the function level and Llama3.1 obtains the highest efficient@1 of 46.51\% at the file level. As for the speedup, GPT-4o achieves the highest speedup of 8.28 at the function level, and Mixtral obtains the highest speedup of 1.43 at the file level.

\finding{1}{DeepSeek V2 Coder and Llama3.1 have the highest probability of generating efficient code solutions with an efficient@1 of 46.97\% and 46.51\%, respectively. GPT-4o and Mixtral generate the most efficient code solutions with a speedup of 8.28 and 1.43, respectively.}

\textbf{Correctness vs. Time Efficiency}
When comparing the correctness and time efficiency of code solutions generated by current LLMs, we find that the best efficient@1 are 46.97\% and 46.51\%, at function level and file level, respectively, which are much lower than the best pass@1 of 79.90\% and 90.78\%. This indicates that almost half of the correctly generated code solutions are sub-optimal since they are less efficient than ground truth solutions. Furthermore, the speedups achieved by most LLMs in file-level code generation are lower than 1.0, and some LLMs even obtain a speedup of lower than 0.1, indicating their generated code solutions are $10\times$ slower than ground truth solutions. This suggests that efficient code generation is a great challenge for current LLMs despite their remarkable performance on correct code generation.

\finding{2}{The performance of current LLMs drops significantly in efficient code generation with the best efficient@1 of 46.97\% and 46.51\% at the function level and file level, compared with that in correct code generation with the best Pass@1 of 79.9\% and 90.78\%. This indicates that the code solutions generated by current LLMs are correct but not time-efficient.}

\textbf{Function-level Code Generation vs. File-level Code Generation.} 
In function-level code generation, we observe that all LLMs achieve a speedup larger than 1.0, indicating that the code solutions generated by current LLMs are more efficient than ground truths. However, only three LLMs achieve a speedup larger than 1.0 in the file-level code generation. This suggests that current LLMs cannot generate faster code solutions than existing solutions in \bench. Besides, the efficient@1 achieved by current LLMs at the function level is also better than that achieved by current LLMs at the file level. For example, the efficient@1 of Phi3 drops by 72.38\% from function-level to file-level code generation. Furthermore, the performance drop from pass@1 to efficient@1 in function-level code generation is 30\%$\sim$45\%, smaller than 40\%$\sim$100\% in file-level code generation. This indicates that current LLMs perform much worse in file-level efficient code generation than function-level efficient code generation.

\finding{3}{Compared with function-level code generation, code solutions generated by current LLMs are less efficient in file-level code generation, evidenced by the significantly lower speedup, lower efficient@, and larger performance drop from pass@1 to efficient@1.}

\textbf{Impacts of Different Model Sizes.}
To study the impacts of different model sizes on the time efficiency of code solutions generated by current LLMs, we observe the changes of pass@1 and efficient@1 from smaller LLMs to larger LLMs. In both function-level and file-level code generation, we find that larger LLMs can generally generate more correct solutions, evidenced by higher Pass@1 obtained by larger LLMs. However, larger LLMs do not always generate more efficient code solutions. For example, in function-level code generation, CodeLlama-34b achieves an efficient@1 of 40.37\%, which is quite close to the efficient@1 of 40.90\% achieved by Llama3-70b and 39.58\% achieved by Llama3.1-405b, but Llama3.1-405b is more than $10\times$ larger than CodeLlama-34b. In file-level code generation, Llama3-70b achieves an efficient@1 of 38.76\%, which is also quite close to the efficient@1 of 41.09\% achieved by DeepSeek V2, and 42.25\% achieved by Deep Seek V2 Coder, but DeepSeek V2 is more than $3\times$ larger than Llama3-70b. 

\finding{4}{Larger LLMs generally perform better in terms of Pass@1 but do not significantly outperform smaller LLMs in terms of efficient@1, indicating larger parameter sizes of current LLMs do not contribute much to efficient code generation.}

In summary, based on the experiment results of 14 popular LLMs on \bench, we study the time efficiency of function-level and file-level code solutions generated by the LLMs in four aspects. We find efficient code generation much more challenging for current LLMs than correct code generation, especially for file-level efficient code generation. We also identify that larger LLMs do not always perform better on efficient code generation.

\section{Implications}

Based on the findings we conclude in Sec.~\ref{sec:eval}, we provide some implications for researchers who build LLMs and practitioners who use LLMs in software development.

\textbf{LLM Researchers.} We identify that there is a large gap between correct code generation and efficient code generation. This indicates that the current LLM-generated code is correct but sub-optimal, and generating efficient code remains a great challenge, especially for file-level code generation. This challenge cannot be effectively mitigated by just increasing the model size of current LLMs. We recommend that LLM researchers consider the code structure and semantics when improving the time efficiency of LLM-generated code. Besides, LLM researchers should also focus more on file-level code generation since current LLMs perform much worse on it than function-level code generation.

\textbf{Software Practitioners.}  As LLMs are gradually adopted in software development in product environments, software practitioners face the problem of choosing LLMs. In function-level and file-level code generation, generally, code solutions generated by DeepSeek V2 Coder and Llama3.1-405b obtain the best time efficiency, respectively. However, we also find that some LLMs with middle sizes, such as Llama3-70b and CodeLlama-34b, achieve competitive performance. We recommend software practitioners adopt middle-sized LLMs to obtain similar performance on efficient code generation with much lower computational costs.
\section{Threats to Validity}\label{sec:discussion}

Our research may face the following threats to the internal and external validity.

\subsection{Threats to Internal Validity}

\textbf{Performance Measurement.} The time efficiency measurement of code solutions generated by LLMs can introduce errors. We propose to use CPU instruction count instead of execution time to improve the stability of measurements. However, there still exist factors such as specific code optimization techniques that introduce measurement errors. To mitigate the threats posed by the errors in time efficiency measurements, we conduct all measurements in dockers~\cite{docker} to ensure that only one single process is running at the same time. Furthermore, we run the measurements for each code solution 12 times and remove the highest and lowest measurements before calculating the average metric. This could further reduce the errors introduced in a single measurement.

\textbf{Baseline Implementation.} Currently, there are no LLM-based stressful test case generation methods that could be compared with \tool, so we modify three correctness test case generation methods as our baselines. However, such modifications may result in performance changes. To improve the validity of baselines, we run them on the most powerful and robust LLM GPT-4o~\cite{gpt4o}. Besides, we ask the baselines to generate 20 stressful test cases once and only choose the best 5 test cases for most evaluations except for accuracy. Therefore, we believe our implementations can represent the best performance of baselines.

\subsection{Threats to External Validity}

\textbf{Adaptation to Different Programming Languages.} While code generation is a general task for all programming languages, we mainly focus on the evaluation of Python code generation in this paper. The code generation performance of LLMs on other programming languages such as C++ and Java may be different from the experiment results we show in Sec.~\ref{sec:eval}, as it could be affected by the syntax and coding styles. This threatens the validity of our experiment results in other programming languages. However, Python is the top 2 most popular programming language at GitHub~\cite{octoverse} and is the major programming language used to build the code generation benchmarks~\cite{humaneval,mbpp,codecontests,apps,effibench,hai24repoexec,repobench,repoeval}. Besides, our stressful test case generation method \tool is language-agnostic and fully based on LLMs to generate stressful test cases, we believe it could be easily extended to build benchmarks for other programming languages.
\section{Related Work}\label{sec:literature}

\subsection{LLMs for Code Generation}
As a critical task to automate the software development process, code generation has drawn a lot of attention in both the academia and industry. At the beginning, encoder-decoder models such as AlphaCode~\cite{codecontests}, CodeT5~\cite{codet5}, CodeRL~\cite{coderl}, CodeT5+~\cite{codet5plus} are directly trained on large code corpus and obtain good performance on code generation. Recently, decoder-only models such as Codex~\cite{codex}, CodeGen~\cite{codegen, codegen2}, InCoder~\cite{incoder}, CodeGeeX~\cite{humanevalx}, SantaCoder~\cite{santacoder}, StarCoder~\cite{starcoder,starcoder2}, WizardCoder~\cite{wizardcoder}, CodeLlama~\cite{codellama}, MagicCoder~\cite{magiccoder}, DeepSeek-Coder~\cite{deepseekcoder} show superior performance than encoder-decoder models on code generation. Besides, some general LLMs trained on multiple types of data, such as Llama3\cite{llama3}, Llama3.1~\cite{llama31}, GPT-3.5~\cite{chatgpt}, GPT-4~\cite{gpt4} also demonstrate competitive or even better performance compared with code LLMs.

\subsection{Code Generation Benchmarks}

\textbf{Correctness Benchmarks.} There are many benchmarks designed for the correctness evaluation of code generated by LLMs. They provide contexts that indicate the functionality of the generated code and several test cases to evaluate the correctness of the generated code. The benchmarks are initially built from scratch by skilled developers and researchers. HumanEval~\cite{humaneval} is a benchmark that contains 164 Python programming problems with function signatures and docstrings. MBPP~\cite{mbpp} is a benchmark consisting of 974 basic Python programming problems with short functionality descriptions. It also provides a sanitized version with verified ground truth solutions that have 427 problems. In order to comprehensively evaluate the performance of LLMs, some benchmarks are built from code competition problems. APPS~\cite{apps} contains 10,000 Python problems with different difficulty levels and diversified ground truth solutions for each problem. Code Contests~\cite{codecontests} is a multi-lingual benchmark built from various competition sources and includes both correct and incorrect human solutions for each problem. Apart from Code Contests, there are also other multi-lingual benchmarks such as xCodeEval~\cite{xcodeeval} and HumanEval-X~\cite{humanevalx}. The above-mentioned benchmarks focus on the evaluation of function-level or file-level code generation. There are some research efforts, such as RepoEval~\cite{repoeval}, RepoBench~\cite{repobench}, SWE-Bench~\cite{swebench}, and CrossCodeEval~\cite{crosscodeeval}, devoted to the evaluation of the repo-level code generation performance. 

\textbf{Time Efficiency Benchmarks.} Despite the well-explored evaluation for the correctness of code generated by LLMs, the time efficiency of code generated by LLMs is under-explored.  Effibench~\cite{effibench} is the first benchmark designed for evaluating the time and memory efficiency of code generation.  It selects efficiency-critical problems tagged ``LeetCode'' and prompts GPT-3.5 to generate test cases with different input sizes and data distribution. However, the problems in this benchmark are too difficult, so most open-source models cannot even generate correct solutions. Besides, it adopts execution time as the performance metric, which is unreliable to distinguish the efficiency of different code solutions. There is also some work~\cite{zapa09accuracy,laaber20dynamic,traini23towards} on traditional performance engineering, but they are not suitable for evaluating random responses from LLMs.

\subsection{LLM-based Test Case Generation}
Apart from the advances in code generation, LLMs have also been demonstrated to improve software testing~\cite{deng23large}. A lot of work has comprehensively evaluated the ability of LLMs on test case generation~\cite{li24large,ouédraogo24large,sami24a,karanjai24harnessing,niels24code}. Most recently, Chen \etal~\cite{chen24chatunitest} propose ChatUniTest, a unit test generation framework
based on LLM by utilizing innovative mechanisms such
as adaptive focal context and generation-validation-repair mechanisms. Liu \etal~\cite{liu24llm} propose a novel LLM-powered test oracle generation approach that combines LLMs and differential testing. Hossain \etal~\cite{hossain24togll} propose TOGLL, a fine-tuned LLM on designed instruction prompts to generate test oracle for Java projects. Wang \etal~\cite{wang24testeval} propose TestEval to generate test cases that cover certain lines, branches, and paths of the code under test. Despite the effectiveness of previous approaches on correctness test case generation, there is no work on stressful test case generation that aims to generate large test inputs to evaluate the time efficiency of the code under test. In this paper, we propose a novel approach \tool to generate stressful test cases for Python projects with high accuracy and coverage.
\section{Conclusion}\label{sec:conclusion}

In this paper, we propose a new benchmark \bench for the time efficiency evaluation of LLM-generated code. To address the challenges of existing correctness code generation benchmarks, we propose a novel stressful test case generation method \tool that incorporates contracts and two test case formats to improve the accuracy. We also introduce a new time efficiency metric \textit{efficient@k} based on CPU instruction count that stably evaluates both the correctness and time efficiency of code. Based on \bench, we evaluate 14 popular LLMs and identify four important findings. We provide implications based on the findings for LLM researchers and software practitioners.

\section{Data Availability}
The code and data of \tool and \bench are available at https://github.com/JohnnyPeng18/Coffe.

\section*{Acknowledgment}
The authors would like to thank the anonymous reviewers who have provided insightful and constructive comments on this paper.
This work is supported by the National Nature Science Foundation of China (No. 62302437).

\newpage
\bibliographystyle{ACM-Reference-Format}
\bibliography{ref}

\end{document}